\documentclass[aps,amssymb,amsmath,pra,twocolumn,showpacs,superscriptaddress,longbibliography]{revtex4-2}

\usepackage{graphicx}
\graphicspath{{figs_main/}{figs_SI/}}

\usepackage{dcolumn}
\usepackage{bm}
\usepackage{subfigure}
\usepackage{color}
\usepackage{soul}

\usepackage{amsfonts,verbatim}

\usepackage[margin=0.6in]{geometry}

\usepackage{tikz}

\usepackage[english]{babel}
\usepackage{dsfont}
\usepackage{latexsym}
\usepackage{float}
\usepackage{afterpage}
\usepackage{enumitem}
\usepackage{mleftright}

\definecolor{lR}{rgb}{1, 0.8, 0.79}

\usepackage{eso-pic, graphicx}

\usepackage{listings}
\usepackage{multirow}
\usepackage{xcolor,colortbl}

\usepackage{bbm}
\usepackage{upgreek}
\usepackage{newtxtext,newtxmath}

\newcommand{\nocontentsline}[3]{}
\newcommand{\tocless}[2]{\bgroup\let\addcontentsline=\nocontentsline#1{#2}\egroup}

\definecolor{Ablue}{rgb}{0.96,0.24,0.00}

\definecolor{Abluetitle}{rgb}{0.,0.24,0.51}

\definecolor{orange}{rgb}{0.96,0.24,0.00}

\definecolor{darkred}{rgb}{0.55, 0.0, 0.0}

\definecolor{darksalmon}{rgb}{0.91, 0.59, 0.48}
\definecolor{maroon}{cmyk}{0,0.87,0.68,0.32}

\setcitestyle{numbers,square,citesep={,\kern-.24em}}

\definecolor{mustard}{rgb}{1.0, 0.86, 0.35}

\addto\captionsenglish{\renewcommand{\figurename}{Fig.}}

\definecolor{Gray}{gray}{0.85}
\definecolor{LightCyan}{rgb}{0.88,1,1}
\newcolumntype{a}{$>${\columncolor{Gray}}c}
\newcolumntype{b}{$>${\columncolor{White}}c}

\usepackage{array}
\newcolumntype{L}[1]{$>${\raggedright\let\newline\\\arraybackslash\hspace{0pt}}m{#1}}
\newcolumntype{C}[1]{$>${\centering\let\newline\\\arraybackslash\hspace{0pt}}m{#1}}
\newcolumntype{R}[1]{$>${\raggedleft\let\newline\\\arraybackslash\hspace{0pt}}m{#1}}

\newcolumntype{P}[1]{>{\centering\arraybackslash}p{#1}}
\newcolumntype{M}[1]{>{\centering\arraybackslash}m{#1}}

\usepackage[colorlinks=true , citecolor=black,urlcolor=blue]{hyperref}

\newcommand{\xt}{\vartheta}

\newcommand{\app}{\approx}

\newcommand{\Cs}{{}^{13}\R{C}}

\newcommand{\xD}{\Delta}

\newcommand{\rt}{\rightarrow}

\newcommand{\beq}{\begin{equation}}
\newcommand{\eeq}{\end{equation}}
                  
\newcommand{\benum}{\begin{enumerate}}
\newcommand{\eenum}{\end{enumerate}}
                    
\newcommand{\bit}{\begin{itemize}}
\newcommand{\eit}{\end{itemize}}

\newcommand{\zhat}{\hat{\T{z}}}

\newcommand{\bea}{\begin{eqnarray}}
\newcommand{\eea}{\end{eqnarray}}

\newcommand{\bev}{\begin{verbatim}}
\newcommand{\eev}{\end{verbatim}}

\newcommand{\T}[1]{\textbf{#1}}
\newcommand{\I}[1]{\textit{#1}}
\newcommand{\R}[1]{\textrm{#1}}

\newcommand{\zfl}[1]{\protect\label{fig:#1}}
\newcommand{\zfr}[1]{\figurename\,\ref{fig:#1}}

\newcommand{\zsl}[1]{\label{sec:#1}}
\newcommand{\zsr}[1]{\!\ref{sec:#1}}

\newcommand{\ba}{\left\{ \begin{array}{lr}}
\newcommand{\ea}{\end{array}\right.}

\newcommand{\blist}[1]{
 \begin{list}{#1}
 \begin{align}
	 arrow
 \end{align}
 $\checkmark\star
  { \setlength{\itemsep}{3pt}
     \setlength{\parsep}{2pt}
     \setlength{\topsep}{3pt}
     \setlength{\partopsep}{0pt}
     \setlength{\leftmargin}{1em}
     \setlength{\labelwidth}{1em}
     \setlength{\labelsep}{0.5em} } }
\newcommand{\elist}{
  \end{list}  }

\DeclareMathSymbol{\vartheta}{\mathalpha}{letters}{"12}
\DeclareMathSymbol{\theta}{\mathalpha}{letters}{"23}
\DeclareMathSymbol{\phi}{\mathalpha}{letters}{"27}
\DeclareMathSymbol{\varphi}{\mathalpha}{letters}{"1E}

\newcommand{\bef}
{
\begin{figure}[htbp]
\centering
}

\newcommand{\eef}{\end{figure}}

\mleftright
\makeatletter
\@addtoreset{section}{part}
\makeatother

\newcommand{\affA}{Department of Chemistry, University of California, Berkeley, Berkeley, CA 94720, USA.}

\newcommand{\affE}{Chemical Sciences Division,  Lawrence Berkeley National Laboratory,  Berkeley, CA 94720, USA.}
\newcommand{\affF}{CIFAR Azrieli Global Scholars Program, 661 University Ave, Toronto, ON M5G 1M1, Canada.}

\begin{document}
\title{Cryogenic field-cycling instrument for optical NMR hyperpolarization studies}
\author{Noella D'Souza}\thanks{These authors contributed equally to this work}\affiliation{\affA}\affiliation{\affE}
\author{Kieren A Harkins}\thanks{These authors contributed equally to this work}\affiliation{\affA}
\author{Cooper Selco}\affiliation{\affA}\affiliation{\affE}
\author{Ushoshi Basumallick}\affiliation{\affA}
\author{Samantha Breuer}\affiliation{\affA}
\author{Zhuorui Zhang}\affiliation{\affA}
\author{Paul Reshetikhin}\affiliation{\affA}
\author{Marcus Ho}\affiliation{\affA}
\author{Aniruddha Nayak}\affiliation{\affA}
\author{Maxwell McAllister}\affiliation{\affA}
\author{Emanuel Druga}\affiliation{\affA}
\author{David Marchiori}\email{dmarchiori@berkeley.edu}\affiliation{\affA}
\author{Ashok Ajoy}\email{ashokaj@berkeley.edu}\affiliation{\affA}\affiliation{\affE}\affiliation{\affF}

\begin{abstract}

Optical dynamic nuclear polarization (DNP) offers an attractive approach to enhancing the sensitivity of nuclear magnetic resonance (NMR) spectroscopy. Efficient, optically-generated electron polarization can be leveraged to operate across a broad range of temperatures and magnetic fields, making it particularly appealing for applications requiring high DNP efficiency or spatial resolution.
While a large class of systems hold promise for optical DNP, many candidates display both variable electron polarizability and electron and nuclear $T_1$ relaxation times as functions of magnetic field and temperature. This necessitates tools capable of studying DNP under diverse experimental conditions.
To address this, we introduce a cryogenic field cycling instrument that facilitates optical DNP studies across a wide range of magnetic fields (10mT–9.4T) and temperatures (${\sim}$10K–300K). Continuous cryogen replenishment enables sustained, long-term operation. Additionally, the system supports the ability to manipulate and probe hyperpolarized nuclear spins via pulse sequences involving millions of RF pulses.
We describe innovations in the device design and demonstrate its operation on a model system of $^{13}$C nuclear spins in diamond polarized through optically pumped nitrogen vacancy (NV) centers. We anticipate the use of the instrument for a broad range of optical DNP systems and studies.

\end{abstract}

\maketitle
\pagebreak

\begin{figure*}[t]
  \centering
  {\includegraphics[width=0.97\textwidth]{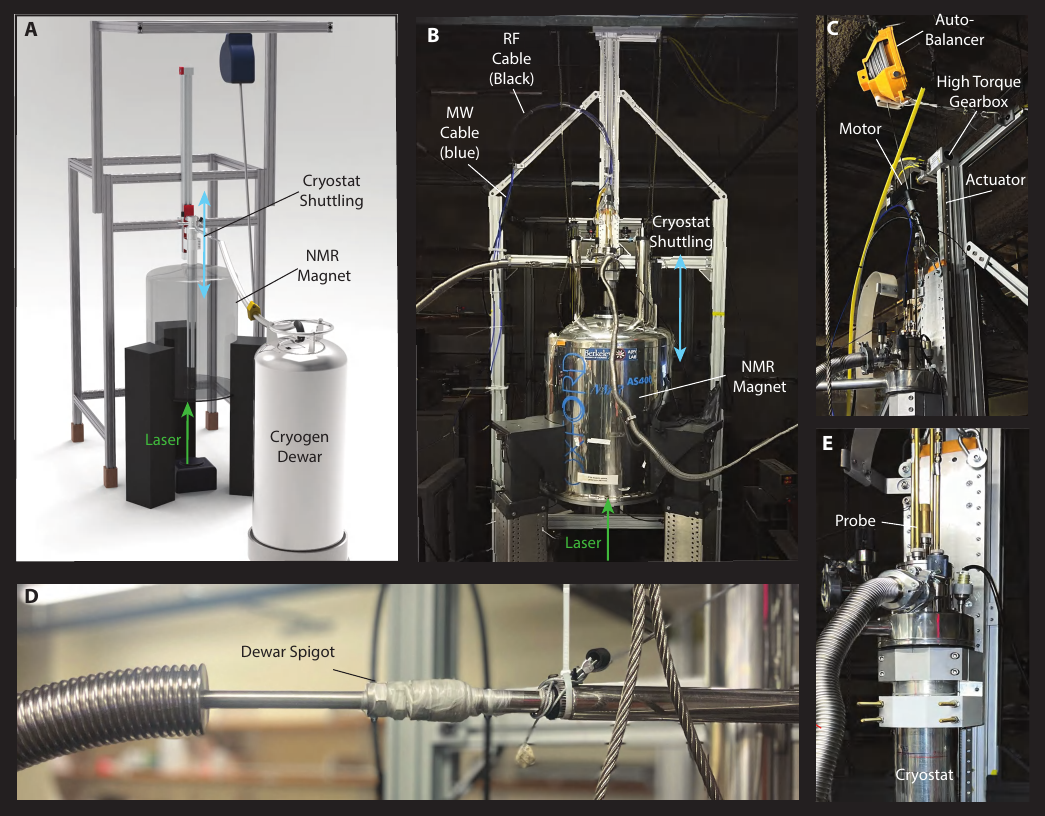}}
  \caption{\T{Overview of instrument.} (A) \I{CAD rendering} illustrates the instrument, comprising a 9.4 T high-field NMR magnet and a 4K-compatible cryostat for sample and NMR probe housing (\zfr{mfig2}). A 532nm laser beam aligns with the cryostat's vertical axis so that the sample can be hyperpolarized either inside or outside the magnet bore. NMR measurements occur at high-field (9.4 T). Continuous cryogen replenishment is sourced from a large dewar (\T{Approach I}) or through a closed-cycle liquid Helium system (\T{Approach II}, see \zfr{mfig2.5}).
(B)  \I{Photograph} of the instrument, mounted on an Oxford NMR magnet.
(C) Close-up shows the motor connected to a high torque gearbox and the belt-driven actuator. A spring-loaded load balancer on the shuttler truck provides extra support to the cryostat against eddy currents during shuttling.
(D) Cryogens are supplied from a non-magnetic dewar (\T{Approach I}) via a Parafilm-sealed transfer line that moves with the cryostat, enabling continuous shuttling. Temperature is regulated by a Mercury ITC (not pictured), and a dual-pump system evacuates the cryostat outer jacket for insulation. An identical strategy can be employed for interfacing to a closed cycle He system (\T{Approach II}).
(E) Cryostat is mechanically mounted on the shuttling truck using oppositely-oriented clamps for vertical alignment with the magnet bore. A custom cap provides access to probe tuning rods.}
\zfl{mfig1}
\end{figure*}

\begin{figure}[t]
  \centering
  {\includegraphics[width=0.49\textwidth]{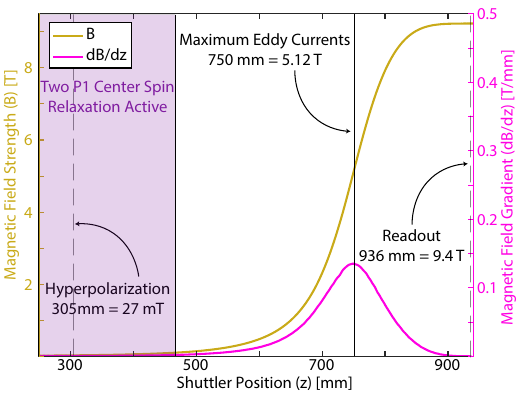}}
  \caption{\T{Field map} indicating variation of the magnetic field with vertical position along the central axis of the magnet. Sweet spot of the magnet is evident at the high-field plateau region (9.4 T). For the experiments on hyperpolarizing $\Cs$ nuclei in diamond, the DNP and readout fields are marked by dashed line. Region where two P1 center-mediated nuclear spin flip relaxation process is active, up to 100mT, is shaded in purple for clarity. Eddy currents are strongest at the maximum field gradient, here at 750 mm (5.1 T).  
}
\zfl{mfig4}
\end{figure}

\section{Introduction}
Dynamic nuclear polarization (DNP)~\cite{Slichter61,Goldman} offers a compelling solution to nuclear magnetic resonance (NMR)’s primary limitation: low sensitivity. By transferring spin polarization from electrons \I{(e)} to nuclei \I{(n)} via DNP, NMR signals can be amplified by more than two orders of magnitude over thermal conditions \cite{Maly08}. This enables signal-enhanced spectroscopy~\cite{Jaroniec04} and expands the potential of nuclear spins, with their inherently long lifetimes, for emerging applications in quantum sensing~\cite{Degen17, Sahin2022} and spintronics~\cite{Reimer10}.

Optically-generated dynamic nuclear polarization (DNP) offers compelling advantages over traditional thermal (Boltzmann) methods~\cite{Henstra88,Tycko98}. Unlike conventional techniques, optical DNP is not limited by thermal electronic polarization, enabling high electron polarization — essential for DNP — to be achieved across a wider range of temperatures and magnetic fields. This flexibility enhances both accessibility and efficiency. Key strengths of optical DNP include: \I{(i)} the possibility of continuous polarization replenishment via optical pumping, \I{(ii)} highly localized polarization with spatial resolution potentially limited only by optical diffraction~\cite{London13}, and \I{(iii)} efficient polarization transfer to nuclei, as electron polarization can be regenerated at rates exceeding $T_{1e}^{-1}$~\cite{Sarkar22}. When paired with operation at lower magnetic fields, where electron transitions lie within low-frequency microwave (MW) ranges accessible by readily available high-power sources, optical DNP can yield significant enhancements to the \I{rates} of nuclear spin polarization. For instance, Ref.~\cite{Sarkar22} demonstrated a ${>}$100-fold increase in the injection rate of $^{13}$C nuclear spins in diamond polarized via Nitrogen Vacancy (NV) centers through a combined use of high-power optical and MW excitation at low fields.

Optical DNP is being actively explored across a wide array of systems, including: \I{(i)} semiconductors such as diamond and SiC~\cite{Fischer13,Alvarez15,Ajoy17}, \I{(ii)} organic molecules hosting triplet-polarizable electrons (e.g. pentacene~\cite{tateishi2014room,hautle2024creating,singh2024room,mena2024room}, porphyrins~\cite{hamachi2021porphyrins}, and derivatives~\cite{sakamoto2023polarizing}) and sensitizer-acceptor pairs via optical CIDNP~\cite{de2023optically}, and \I{(iii)} molecular complexes with rare earth inorganic (REI) centers~\cite{bayliss2020optically,Laorenza21}. The latter two classes are particularly interesting due to their synthetic tunability. Chemical modification can allow precise control over properties such as polarization agent spacing and topology, potentially enabling the development of custom-designed, bottom-up engineered molecular frameworks optimized to host polarization sources~\cite{zadrozny2017porous}. 

However, there is a large variation in electron polarizability and electron and nuclear $T_1$ relaxation times in these materials as functions of magnetic field ($B$) and temperature ($T$). This motivates the need for tools to be able to probe nuclear hyperpolarization under diverse $B$ and $T$ conditions. To address this, we report the development of a novel instrument enabling optical DNP studies across a broad range of temperatures (from room temperature to $~\sim$10K) and magnetic fields (10mT to 9.4T). The device also facilitates the interrogation and manipulation of hyperpolarized nuclear spins under millions of applied RF pulses.

The core methodology, “cryogenic field cycling”, involves shuttling a sample within a cryostat between low-field polarization and high-field NMR detection regions. Our design builds on prior advances in field cycling~\cite{Ajoy_widecycle_2019,Kiryutin16,Zhukov18,Charlier13,Hall20}, extending the operational temperature range to $~\sim$10K and enabling spin control with ${>}$8 million RF pulses, albeit with trade-offs in shuttling speeds due to eddy currents from moving a cryostat into a magnet. We detail the design, construction, and operation of this instrument. Results are demonstrated on a model system of $^{13}$C nuclear spins in diamond, but the device itself is applicable to a wider range of materials and DNP applications.

\section{Cryogenic field cycling: Design and construction}
\zsl{design}
\subsection{Field cycling a Cryostat: Principle and Construction}

\zfr{mfig1}A-B provides an overview of the instrument. Central to its operation is the mechanical shuttling of a cryostat (Oxford SpectrostatNMR) (\zfr{mfig1}C) under continuous flow cryogenic cooling (\zfr{mfig1}D). Inside the cryostat, the sample of interest is subject to MW-driven optical DNP at low magnetic fields and then shuttled into high-field (9.4 T) for NMR readout (\zfr{mfig1}E). A dynamic-load pulley system aids with shuttling the 25 lb cryostat in opposition to the force of eddy currents. While the field-cycling arrangement here builds on the system described in Ref.~\cite{Ajoy_widecycle_2019}, this system introduces operation down to cryogenic temperatures $~\sim$10K. This is enabled by two innovations, described below, which are focused on direct shuttling of an entire cryostat, accounting for the presence of strong eddy current forces, and continuous replenishment of cryogen to maintain sample temperature during shuttling.

An aluminum extrusion frame, built around the NMR magnet, hosts a belt-driven actuator (Parker HMRB08) upon which the cryostat is mounted (\zfr{mfig1}A-B). The actuator offers excellent (50$\mu$m) spatial precision, and a motor (ACS) fitted with a high-torque gearbox facilitates actuator operation by enhancing load-carrying capacity (\zfr{mfig1}C). The actuator carries a movable stage (“truck”) on which the cryostat is secured by two large, custom-designed clamps (\zfr{mfig1}E). The alignment of the cryostat in the magnet bore is maintained ${>}$0.5° over the entire distance of travel; this is necessitated by the tight clearance (${<}$2mm) of the cryostat outer walls to the magnet bore.

In physically attempting to shuttle the entire cryostat assembly, eddy current forces present a significant challenge~\cite{Conradi19}. They oppose the shuttling motion and vary in magnitude proportionally to the gradient of the magnetic field encountered. \zfr{mfig4} illustrates the magnetic field strength and field gradient profile, relative to the sample's position. Notably, the maximum force arises 186 mm before the sample achieves the NMR magnet's maximum field, or "sweet spot."

To address the effects of eddy currents, we employ a two-pronged strategy. Firstly, the mechanical design of the probe is simplified by using the stainless steel cryostat body itself as an RF shield (\zfr{mfig2}B). This approach eliminates the need for a separate shield and reduces the weight of the NMR probe. Stainless steel generates lower eddy current forces compared to higher conductivity materials like copper. Secondly, we incorporate a high-torque gearbox, adding an extra 10 lb. of torque to the shuttling motor to manage the increased load from eddy currents. Lastly, we employ a spring load balancer pulley system (\zfr{mfig1}C) coupled to the shuttling truck as a dynamically variable counterweight, providing vertical force to the truck to counterbalance any additional forces beyond a predetermined set point (33lb.). As a result, eddy current forces encountered during motion are effectively compensated for, ensuring that these loads are not transferred to the motor or the actuator belt.

In our experimental setup, the magnetic field experienced by the sample is controlled by adjusting the vertical position of the cryostat within the magnet’s fringe field. The field profile during the sample’s journey is shown in \zfr{mfig4}, with the transition from 27 mT to 9.4 T taking ${\sim}$91s.

This relatively long shuttling time represents a design tradeoff inherent to the shuttled cryostat configuration. While the cryostat enables low-temperature operation, it significantly increases shuttling duration compared to similar room-temperature devices (e.g., 0.6s in Ref.~\cite{Ajoy_widecycle_2019}). However, this limitation is not prohibitive for the majority of target samples for two key reasons.

First, at cryogenic temperatures and high magnetic fields, nuclear spin relaxation times ($T_1$) for hyperpolarized nuclear spins are inherently long, exceeding the shuttling duration. Second, $T_1$ is strongly field-dependent and typically short only at low fields~\cite{Ajoy19relax}. This region is marked for case of $\Cs$ nuclei in diamond in \zfr{mfig4}. The most critical time scale, therefore, is the duration required to traverse the low-field regime (${<}$1 T), where shorter $T_1$ values are expected. This traversal is completed in ${<}$3s (\zfr{mfig4}), mitigating concerns but nonetheless setting a practical boundary on the types of materials currently suitable on the instrument.

This reasoning aligns with findings in Ref.~\cite{Ajoy19relax}, which examined nuclear $T_1$ relaxation in a model electron-rich solid. The study revealed that $T_1$ increases significantly when the nuclear Larmor frequency surpasses the electron paramagnetic resonance (EPR) linewidth, as two-electron spin flip processes—while dominant at low fields—are suppressed at higher fields. 

\subsection{Replenishable cryogen delivery compatible with field-cycling}
The instrument can maintain cryogenic sample conditions over extended periods, including continuous field-cycling operation lasting several days. This capability advances over earlier methods that applied cryogens only intermittently ~\cite{Kiryutin16}.

We present two different designs for cryogen delivery. \T{Approach I}, optimized for the 77K to room temperature (RT) range, employs a flow-through system fed by a liquid nitrogen (LN) dewar. In this configuration, the cryogen is not recovered, but a typical 100L dewar supports continuous operation for approximately two weeks. Refills can be performed with a secondary dewar while maintaing cryogenic conditions, ensuring uninterrupted operation. \T{Approach II}, designed for lower temperatures (reliably down to 10K, and in principle down to ${\app}$4K), utilizes a closed-loop liquid helium (LHe) system. This configuration offers extended low-temperature operation via cryogen recycling.

\subsubsection{\T{Approach I:} Cryogen Delivery System for 77K to Room Temperature}

We now describe \T{Approach I} that allows operating temperature from 77K-RT. Cryogen delivery to the cryostat is achieved through a horizontal plunging spigot (as shown zoomed in \zfr{mfig1}D).  The cryostat’s outer layer is maintained under vacuum by a dual pump system, comprising a Pfeiffer ONF 016 Vacuum Duo Line roughing pump and a Pfeiffer D35614 HiCUBE 80 Eco Vacuum TurboPump. This setup achieves an ultralow pressure of approximately $10^{-7}-10^{-8}$ bar, ensuring optimal thermal insulation.

The mechanical design, featuring a motor and spring-loaded pulley system, stabilizes the sidearm and transfer line throughout the field-cycling transit (\zfr{mfig1}C). Adequate slack is maintained to keep the spigot near-horizontal (\zfr{mfig1}B,D), minimizing the risk of leaks. The entire spigot moves with the cryostat, enabling prolonged low-temperature operation. Temperature regulation is achieved using an Oxford Instruments ITC503 controller paired with a variable-output heater, maintaining the desired temperature setting.

The end of the transfer line is connected to a LN dewar (as depicted in \zfr{mfig1}A),  facilitating a continuous nitrogen vapor flow to maintain sample temperatures at or above ${\app}$80 K. This system has been tested for extended durability, sustaining continuous shuttling every few minutes for over two months without manual intervention.

\subsubsection{\T{Approach II:} Closed-Cycle Liquid Helium (LHe) System for Low Temperatures}

\T{Approach II} enables access to temperatures reliably down to $~\sim$10K using a closed-cycle liquid helium (LHe) system (ColdEdge Stinger). This system’s primary advantage lies in LHe recovery, which eliminates the need to periodically refill cryogen. This removes an otherwise prohibitive cost for long-term experiments. 

The ColdEdge Stinger employs a Gifford-McMahon (GM) refrigeration cycle to cool the cryocooler down to 4K. Room-temperature helium gas is then circulated over the cryocooler and through the closed-cycle cryostat, progressively cooling the gas and the sample to temperatures near 4K.

As illustrated in \zfr{mfig2.5}A-B, the system incorporates a 4K-compatible cryostat attached to the Stinger.  A similar concept is employed as in \T{Approach I}; the cryostat is shuttled into the magnet for field cycling. The spigot design of \zfr{mfig1}D is replaced with a transfer line that interfaces directly with the Stinger system (see \zfr{mfig2.5}B). 

\zfr{mfig2.5}A shows the assembled system. On top is the Stinger cryocooler which is mounted to an aluminum extrusion (80/20) frame ${\app}$7 ft off the ground  so that the transfer line can reach the cryostat during shuttling. The GM refrigeration cycle is powered by pre-cooled, high-pressure helium gas supplied by the helium compressor (Sumitomo, F-70) located at the base of the setup. Vacuum is pulled on the vacuum jacket of the cryocooler simultaneously to the vacuum jacket of the cryostat, using the dual vacuum pump system described previously. 

In addition to the cryocooler, there is also an auxiliary helium circuit which circulates helium gas using the top compressor in \zfr{mfig2.5} (Sumitomo, CKW-21). This compressor pushes room temperature helium gas over the cryocooler and through the cryostat, where it is connected to aflexible transfer line \zfr{mfig2.5}B which moves synchronously with the cryostat during shuttling and allows sustained cooling operation even during cryostat motion.

To ensure proper compatibility, we employ a custom cryostat from Coldedge, as is shown in \zfr{mfig2.5}C.  It features three distinct chambers: the outermost vacuum-insulated chamber, a middle chamber for cold helium gas circulation, and the innermost chamber housing the sample. The innermost chamber includes an external valve for easy access and is continuously purged with helium gas during cooldown. This purging enhances heat exchange between the sample and cryogen while maintaining constant pressure during cooling. Each chamber is sealed with quartz windows to enable laser access for hyperpolarization experiments.

The cryocooler can cool to ${\sim}$5K in 75 mins. Cooling the sample well to below 10K requires approximately 3 hours, with further stabilization to 4.3K taking around 18 hours when the sample well is empty. While sample well temperatures of ~${\sim}$8K have been achieved with the probe incorporated into the sample well; typically temperatures of $~\sim$10 K are achieved for experiments that include shuttling. Compared to \T{Approach I}, this closed-loop configuration demands higher vacuum integrity to ensure cryogen flow, which is critical for reliable and sustained operation.

 \begin{figure}[t]
  \centering
  {\includegraphics[width=0.49\textwidth]{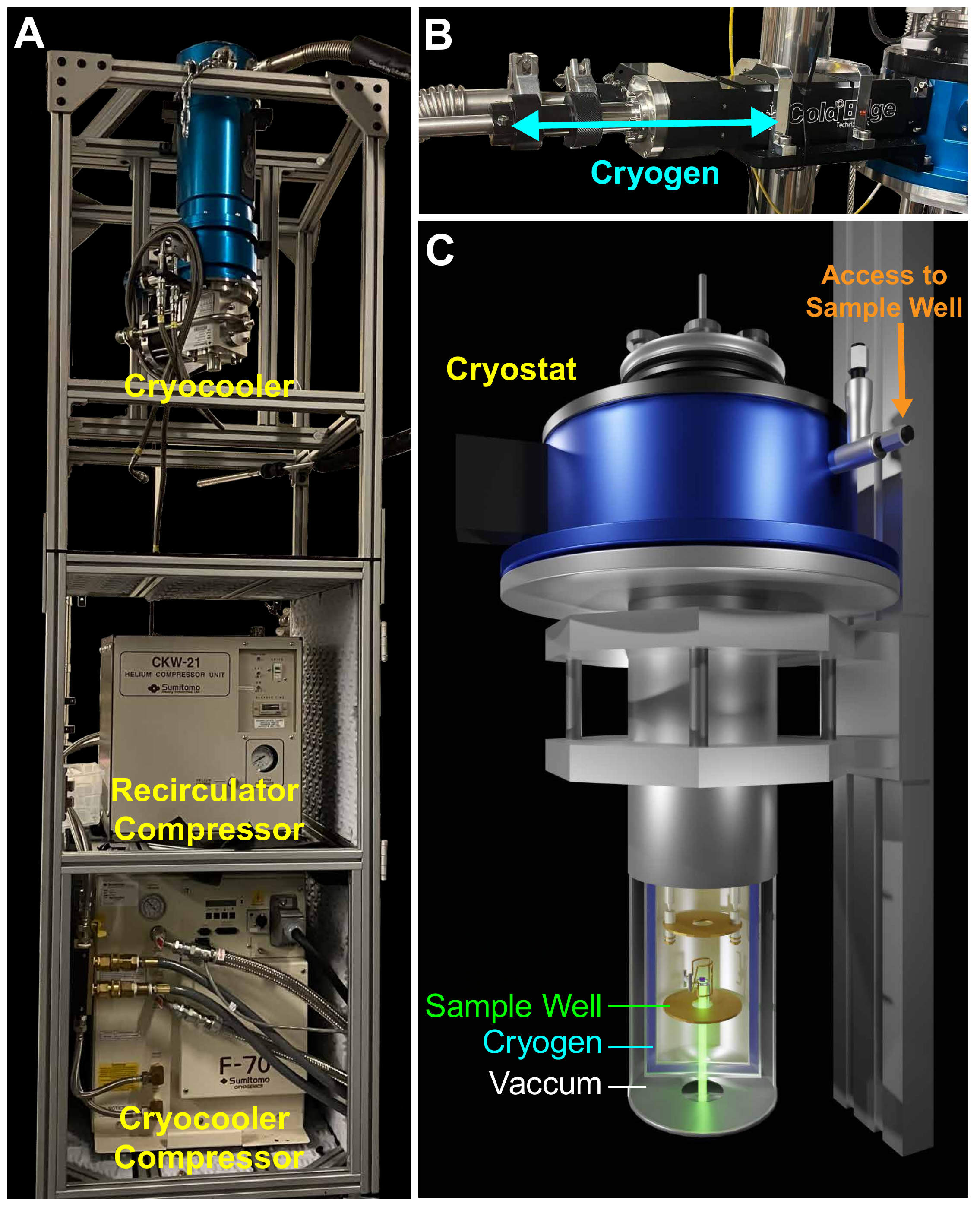}}
  \caption{\T{Closed-cycle liquid helium system} allowing operation $~\sim$10K. (A) \I{Mounted system} displaying the compressors and cryocooler. The top displays the Coldedge Stinger cryocooler (blue) which is cooled using the Sumitomo helium cryocooler compressor on the bottom. Above it, the recirculator He compressor moves He gas through the cryocooler and cryostat. (B) \I{Transfer line} connects the helium gas flowing over the cryocooler to the cryostat and is compatible with continuous shuttling of the cryostat. (C) \I{CAD rendering of the cryostat} with mounted probe. The length of cryostat is not to scale. The cryostat has three separate chambers: the outer-most chamber is a vacuum jacket, the central chamber is where the cryogen flows, and the inner-most chamber is where the sample is placed. The valve on the outside of the cryostat can be used to purge the sample chamber with helium gas. Each chamber is sealed with a quartz window, permitting laser access for hyperpolarization.
}
\zfl{mfig2.5}
\end{figure}

\subsection{Optical excitation for nuclear hyperpolarization}

The instrument enables optical illumination across a broad magnetic field range (10 mT to 9.4 T), an advance over previous work~\cite{Ajoy_widecycle_2019} that restricted optical DNP to a single illumination field. This flexibility allows in-depth exploration of optically mediated DNP mechanisms across diverse field regimes: \I{(i)} low fields (${<}$0.1 T), where electronic hyperfine interactions dominate nuclear Larmor frequencies~\cite{Zangara18, Pillai21}; \I{(ii)} intermediate fields (e.g., X-band), routinely used for integrated solid-effect mechanisms and related processes~\cite{Henstra90, tateishi2014room}; and \I{(iii)} high fields (${>}$7 T), where emerging all-optical nuclear hyperpolarization strategies are gaining renewed interest~\cite{King10, Concilio21, Kuprov22}.

We have implemented two strategies for optical illumination, enabled by a custom-designed probe (\zfr{mfig2}). The first involves directing a high-power free-space coupled laser beam from below the cryostat through a bottom-mounted optical window in the cryostat to access the sample housed in the NMR coil (\zfr{mfig2}G). A 532 nm continuous-wave laser (Coherent 1169898) is guided into the instrument’s bore via a 45-degree mirror (\zfr{mfig3}). To ensure precise delivery, the beam travels a 4-foot path to the cryostat, aligned using piezo-driven mirrors. The laser is stabilized at 19$^{\circ}$C using a Lytron RC006G03BB2M007 chiller.

A camera with a long-pass filter is integrated at the top of the cryostat to capture the sample’s fluorescence, aiding in precise optical alignment. This configuration supports fluorescence-based optical detection, such as optically detected magnetic resonance (ODMR) measurements. ODMR readout can assist in determining microwave frequencies necessary for hyperpolarization excitation.

In the second approach, laser excitation is delivered via a multimode fiber-coupled laser diode source mounted directly on the shuttling track, enabling simple co-shuttling with the cryostat. A key advantage of this method is the ability to position the fiber end in close proximity to the sample, simplifying alignment. We find that the multimode fiber delivers consistent power output, remaining stable down to temperatures of $~\sim$10K.

\begin{figure}[t]
  \centering
  {\includegraphics[width=0.49\textwidth]{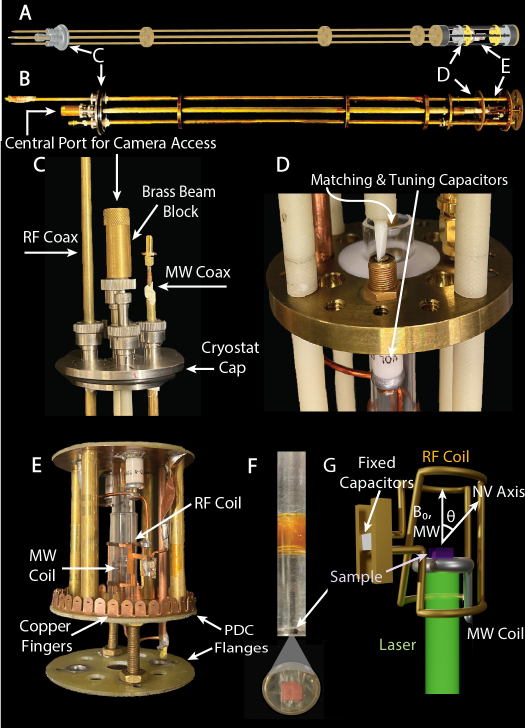}}
  \caption{\T{DNP/NMR Probe.} (A) \I{CAD rendering} illustrates probe design and corresponding figure insets. 
(B) \I{Photograph of probe} with length 39in. from the top of the cap to the bottom of the last support. Probe fits snugly into cryostat, which acts as RF shield during NMR measurements. Copper ``fingers" maintain contact with the cryostat. Support for RF and MW coax along with tuning rods is provided by brass flanges, with PDC flanges located near the coil.
(C) \I{Probe top cap} seals the cryostat.
Custom Swagelok fittings with o-rings enable NMR frequency tuning and impedance-matching to 50 Ohm while maintaining vacuum and cryogenic conditions. The same fittings are used for MW and RF coaxes. A central port, closed with a brass insert when not in use, enables camera access for laser alignment. 
(D) \I{Tuning setup} consists of two tunable capacitors adjusted by ceramic screwdriver rod ends. Tuning rod stability is provided by a self-locking mechanism.
(E) \I{MW and NMR coils.} The MW loop is compactly centered within and perpendicularly oriented to the RF coil shown. (F) \I{Sample mounting} is achieved using a glass insert notched for the sample size.
(G) \I{Schematic of sample orientation} with respect to magnetic field and coil geometry.
  }
\zfl{mfig2}
\end{figure}

\subsection{DNP-NMR probe with shuttling}
The NMR/DNP probe is mounted on the cryostat and shuttled along with it. A schematic of the probe is shown in \zfr{mfig2}A while photographs of the probe are shown in \zfr{mfig2}B-F. This dual-purpose probe supports microwave (MW) excitation for hyperpolarization at low fields and NMR detection at high fields. Both MW and RF coils are integrated within the probe to ensure seamless operation throughout the entire experimental sequence. The probe design also incorporates features that minimize noise pickup during the cryostat’s shuttling motion. 

\T{Construction:} As illustrated in \zfr{mfig2}A-B, the probe is constructed from brass, featuring machined ribs and flanges that accommodate tuning rods, RF and MW coaxial cables, and a central aperture for laser beam passage (\zfr{mfig2}C). Coaxial cables also contribute to the probe's structural stability. Probe matching and tuning can be performed outside of the cryostat (\zfr{mfig2}D) with long rods reaching the RF circuit at the bottom of the probe (\zfr{mfig2}E). The sample is securely held in a quartz tube, with its orientation and position maintained by a notched quartz rod (\zfr{mfig2}F). This aids in the optical detection of fluorescence for laser alignment. For \T{Approach II}, \zfr{mfig2.5}C provides a schematic of the probe integrated within the 4K-compatible cryostat. 

\T{NMR Coil:} The MW and RF coils are designed with compact geometries to minimize mutual interference while keeping the central axis unobstructed for laser illumination (\zfr{mfig2}E). The saddle-shaped RF coil is laser-cut from a 0.7 mm copper sheet using a high-power (60 W) CO$_2$ laser. With an internal diameter of 12 mm, it accommodates both the MW loop and the sample (\zfr{mfig2}E-G). The RF coil is self-supporting in design and requires no external “former” to maintain its shape. Tuning to the nuclear Larmor frequency is achieved using a combination of fixed and tunable non-magnetic capacitors (\zfr{mfig2}D). For example, a 100 MHz system, required for $^{13}$C nuclei at 9.4 T, corresponds to 10 pF fixed and 0.8-10 pF tunable capacitors.

A notable feature of the probe is its ability to deliver 1–10 million pulses to the nuclear spins, with interrogation occurring in intervals between pulses  (\zfr{mfig5}). This requires rigid mounting of the coil and associated circuit to avoid artifacts, as any vibration in the coil wires can directly affect the NMR signal. 

The temperature conditions for \T{Approach I} are less stringent, but the same high-power RF pulses used at temperatures ${\sim}10$K in \T{Approach II} lead to probe arcing and present an operational challenge.  Arcing is largely mitigated by using a helium gas environment in the sample chamber instead of pure vacuum, consistent with Paschen’s Law~\cite{Babich_PaschensLaw}. Data collected at 10 K (\zfr{mfig17new}) using PTFE capacitors (Voltronics AT10-4) demonstrate low arcing, even though these capacitors are not explicitly rated for cryogenic conditions. Alternate capacitor designs, such as those from Ref.~\cite{Tagami_CryoProbe}, have previously shown promising performance in cryogenic environments and could further enhance system reliability.

\T{Tuning and Matching RF Circuitry:} Matching and tuning the NMR circuitry at cryogenic temperatures presents another challenge, as adjustments made at room temperature are incompatible at low temperature. Our solution involves externally-adjustable capacitors, accessible even when the probe is under high vacuum and cryogenic conditions (\zfr{mfig2}C). O-rings fit into ports on the cryostat cap, permitting rotation of the long (3.5 ft) matching and tuning rods while maintaining vacuum. The rods terminate on ceramic knife edges and have a self-locking system to ensure engagement with the tunable capacitors (highlighted in \zfr{mfig2}D).

To ensure good electrical shielding and, therefore, enable high SNR NMR measurements, laser-cut copper “fingers” (shown in \zfr{mfig2}B, E) are attached to the brass flanges in order to provide strong electrical contact with the walls of the cryostat. The stainless steel body of the cryostat acts here as the RF shield.

\T{DNP Coil:} MW excitation is generated using a loop coil oriented perpendicularly to the RF coil, a configuration that minimizes interference and field inhomogeneity during RF detection. This is shown in \zfr{mfig2}E,G. For low-field hyperpolarization of $^{13}$C nuclei, the electronic quantization axis of NV centers is defined by the intrinsic zero-field splitting. The magnetic field component perpendicular to this axis drives dynamic nuclear polarization (DNP)~\cite{Pillai21} (\zfr{mfig2}G).

\T{Shuttling Compatibility:} Flexible, shielded RF and MW coaxial cables from the NMR pre-amplifier and MW amplifiers remain connected to the probe throughout shuttling and are supported by cable guides.  Guides ensure stable operation and minimal noise pickup during NMR acquisition, even after multiple shuttling cycles. The cryostat, with all coaxial cables attached, can be shuttled without compromising integrity.

\begin{figure}[t]
  \centering
  {\includegraphics[width=0.49\textwidth]{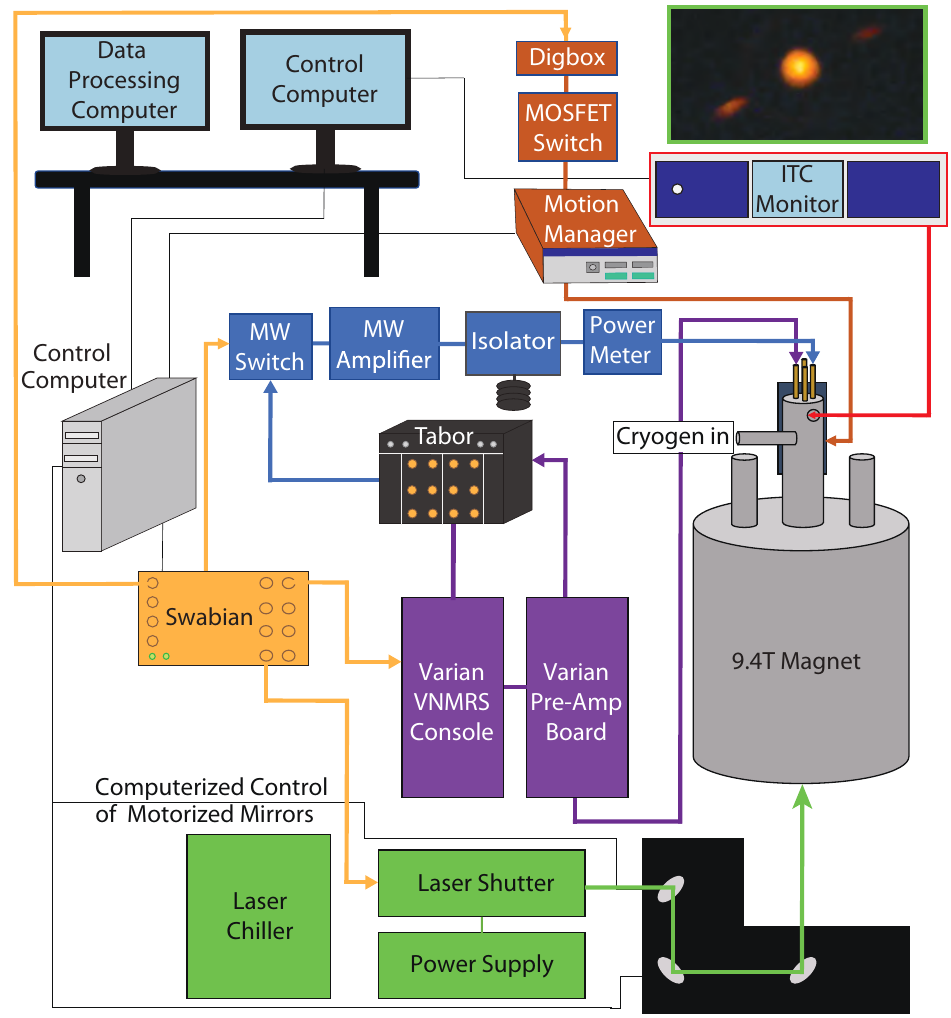}}
  \caption{\T{Control and synchronization of instrument.} (Yellow) \I{Synchronization:} TTL triggers output by the Swabian pattern generator synchronize the timing of the four main components of the setup: laser, MW, shuttling, and RF. (Green) \I{Optics:} Laser beam is directed through the bore of the magnet onto the sample. The sample is aligned by observing the sample fluorescence on a camera (top right inset) and remotely adjusting th beam path through motorized Zaber mirrors to produce optimal sample fluorescence.
(Blue) \I{Microwave}: MW output by the Tabor Proteus AWT is applied to the sample when the Swabian triggers the MW switch on. An isolator protects the amplifier from reflected power. MW power is tracked with a power meter.
(Orange) \I{Shuttling}: After optical hyperpolarization, the Swabian trigger is stepped up to 5V via the Digbox. This 5V trigger allows the MOSFET switch to connect the 24V circuit and initiate the Motion Manager for cryostat shuttling.
(Purple) \I{Radio-Frequency}: RF pulses, synthesized and amplified in a Varian VNMRS console, are applied to the sample through the RF coil. RF signal is preamplified using the Varian receiver but digitized on the Tabor AWT. 
Temperature control is maintained using a variable-output heater monitored with the Mercury ITC. Temperature information is accessed from the control computer.
}
\zfl{mfig3}
\end{figure}

\subsection{Device synchronization and control}
We now detail the control and synchronization mechanism of the instrument components, illustrated in \zfr{mfig3}, which enable automated operation of the instrument over extended, multi-day experiments. The primary synchronization is provided by a Swabian pulse streamer with 1ns timing resolution. The instrument's belt-driven actuator is managed by an ACS controller. Motion triggers received from the Swabian device ensure precise timing. For safety reasons, the actuator is set to operate at a trigger voltage of 24V. This is created by using a MOSFET to switch a 24V power supply, elevating the Swabian device's trigger to the required voltage (\zfr{mfig3}  \I{Orange}).

The laser system is equipped with a high-speed Thorlabs shutter for effective gating (\zfr{mfig3} \I{Green}) of optical access to the sample and synchronized laser application with MW pulses for hyperpolarization. Motorized Zaber mirrors, used for optical alignment and connected via USB, are adjustable during the shuttling process. This capability eliminates the need for the laser to maintain alignment at both low and high-field locations simultaneously.

The Tabor Proteus AWT device is utilized for generating MW pulses for hyperpolarization (\zfr{mfig3} \I{Blue}) and receiving NMR signals (\zfr{mfig3} \I{Purple}), an extension of the design previously outlined in Ref.~\cite{Beatrez21}. NMR pulse generation is output from a Varian console, with pulse amplification facilitated by a Herley amplifier. The Varian device heterodynes frequencies to 20MHz, which are digitized at 1GS/s, equivalent to sampling every 1ns. In experiments involving multiple pulses (discussed in Sec.~\zsr{spinlock}), digitization is performed during windows between pulses. Data in each interval undergoes Fourier Transform (FT) analysis to extract the amplitude and phase of the Larmor component. The Tabor device, equipped with a large (16GB) onboard memory, is particularly suited for recording the precession signal over millions of pulses. Data transfer from the instrument to a computer is accomplished via a high-speed PCIe bus. Theoretically, the Tabor device has the potential to serve as the foundation for an entire NMR spectrometer, a concept we aim to detail in a forthcoming manuscript.




\begin{figure*}[t]
  \centering
  {\includegraphics[width=0.97\textwidth]{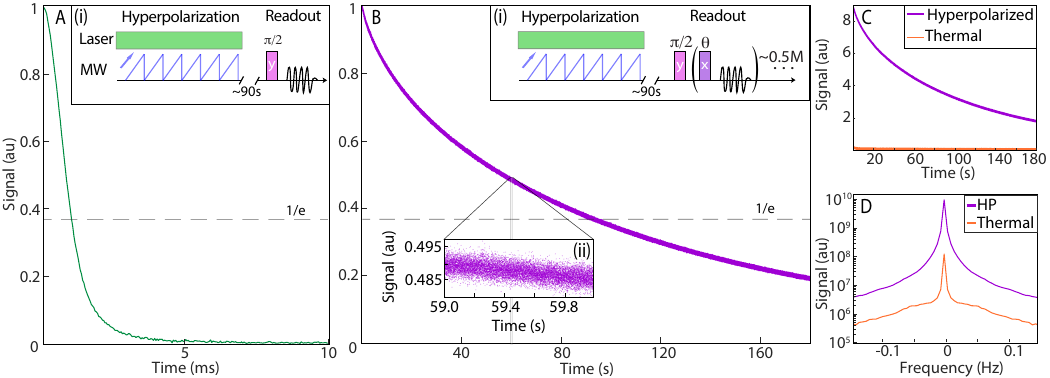}}
  \caption{\T{Probing hyperpolarized $\Cs$ nuclei at 100K.}
(A) FID and (B) pulsed spinlock (SL) traces for $\Cs$ nuclei hyperpolarized via optical DNP at 100K. (i) \I{Inset:} Experimental scheme for DNP followed by an FID readout. Pulsed spin-lock readout involves a pulse train (B(i)) with a 50$\mu$s interpulse separation, and spin interrogation in windows between pulses. (B) Pulsed spinlock data shows ${>}$10,000 fold extension in lifetime, with ${>}$3M points in the trace. Inset shows a 1s window at 59s, and highlights the rapid data sampling rate.
(C) Comparison of thermal (orange) and hyperpolarized (purple) SL signals at 100K clearly demonstrates the SNR improvement from hyperpolarization. For the former, the diamond sample thermalizes in the magnet for 10hr before measurement. The latter involves laser illumination for 60s.
(D) Effective NMR spectra derived from SL time traces in (C), obtained via FT, and displayed on a logarithmic scale. Results indicate a (${>}$100-fold) increase in SNR with DNP.}
\zfl{mfig5}
\end{figure*}

\section{Experiments carried out at cryogenic conditions}
\zsl{prelim}
In this section, we present optical hyperpolarization experiments conducted under cryogenic conditions using a ${\sim}$1ppm NV-center doped single-crystal diamond as a model system. The NV electron is used to optically hyperpolarize surrounding $\Cs$ nuclei. Our choice of this system stems from its well-characterized behavior at room temperature, allowing us to contrast with the less explored spin dynamics at low temperatures, providing insights into temperature-dependent phenomena. Practically, hyperpolarized $\Cs$ nuclear spins in NV-diamond are of considerable interest in applications like quantum sensing and quantum memories.

The primary focus here is to illustrate the capabilities of our instrumentation in facilitating these measurements. Detailed analysis of the underlying physics will be addressed in future publications.

\begin{figure}[t]
  \centering
  {\includegraphics[width=0.49\textwidth]{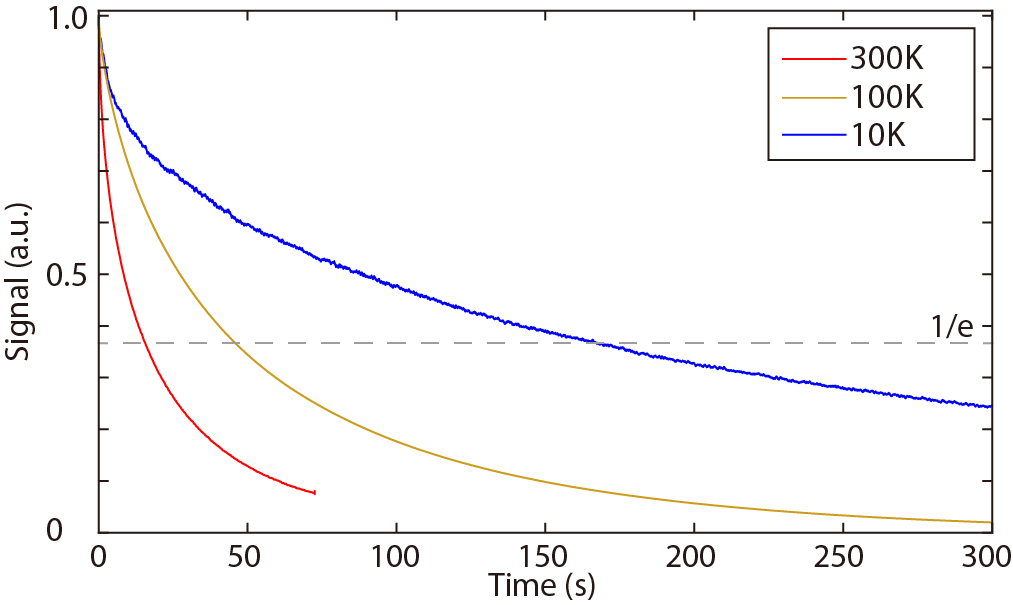}}
  \caption{\T{SL measurements at different temperatures.} Three SL measurements taken at RT, 100K, and 10K. Data at 100K and 10K is taken over a 5min. period involving ${>}4$M pulses. All data is smoothed using a moving boxcar average. Dashed line marks $1/e$ intercept. Data reveals a significant lengthening of $T_2'$ times with decreasing temperature. 
}
\zfl{mfig17new}
\end{figure}

\subsection{Spin-lock multipulse experiments at variable temperatures}
\zsl{spinlock}
High-field NMR measurements offer several advantages over their low-field counterparts, including higher SNR due to lower $1/f$ noise at higher frequencies, extended nuclear lifetimes, and support for a greater degree of spin control. The latter allows us to carry out multipulse experiments with the windowed acquisition, spanning millions of pulses.

First, \zfr{mfig5}A displays the FID of hyperpolarized $\Cs$ spins at 100K. It is characterized by rapid decay primarily governed by internuclear dipolar interactions. In \zfr{mfig5}B, we employ a windowed spin lock (SL) sequence (\zfr{mfig5}B.i). It consists of a series of pulses with 75$\mu$s between the beginning of successive pulses; inductive acquisition of nuclear precession is carried out between pulses, during which the Herley amplifier is blanked.  In \zfr{mfig5}B the data points represent the Fourier intensity at the heterodyned Larmor frequency after every pulse, with the curve comprising over 10M points. There is a significant increase in lifetime, here $T_2'$=93.5s, over 10,000-fold boosted over $T_2^*$=1.1ms. This significantly improves the detection SNR, with a ${>}$200-fold increase in this particular case. 

The SL sequence offers an effective method for evaluating the impact of DNP. For instance, \zfr{mfig5}C compares SL signal traces without (orange) and with (purple) hyperpolarization (HP) at 100K. For thermal measurements, spins thermalize in the magnet for ${>}$10h prior to the experiment. The SNR increase due to ODNP is apparent. To quantify this, we perform an FT on each trace, analyzing the effective line-narrowed NMR spectrum. The results, shown in \zfr{mfig5}D on a logarithmic scale, highlight a ${>}$100-fold gain in SNR from HP. The narrow linewidth observed reflects the long $T_2'$ relaxation time. Given that the DNP process takes only 60s, the acceleration in terms of time efficiency (including the ratio of high and low field $T_1$) is significant, over $10^8$-fold.

\zfr{mfig17new} shows the variation of SL decay profiles at varying temperatures, from RT down to 10 K. An advantage over comparative $T_1$ measurements is the ability to sample the $T_2'$ profiles at an exceptionally high density of points, capturing one data point per pulse). This granularity, capturing  ${>}4$M data points per curve, allows for a deeper understanding of the decay profile, extending beyond just the $1/e$ intercepts (dashed line in \zfr{mfig17new}). The results clearly show an increase in $T_2’$ values as the temperature decreases. The high number of data points facilitates new analytical approaches to study this increase, such as the use of Laplace inversion to examine the constituent decay components~\cite{song2005determining,harkins2024anomalously}. 

We hypothesize that this enhancement arises from the quenching of electron-mediated (NV and P1) noise experienced by nuclear spins. This quenching likely results from an increase in the $T_{1e}$ relaxation times of the electron spins at lower temperatures~\cite{Jarmola12}, and greater electronic polarization at high fields and low temperatures~\cite{Takahashi08}. A more detailed analysis using Laplace inversion of the decay curves could provide deeper insights into individual decay channels and their temperature dependence~\cite{harkins2024anomalously}, which we plan to explore in future work.

\begin{figure}[t]
  \centering
  {\includegraphics[width=0.49\textwidth]{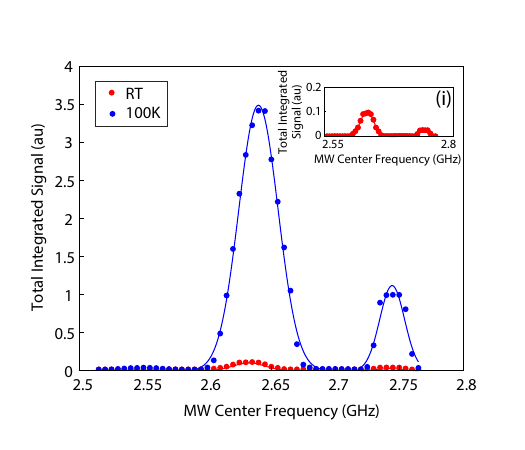}}
  \caption{\T{EPR mapping of NV electron via DNP.} Panel shows DNP-mediated EPR spectra of NV center at two temperatures (RT and 100K). Each point represents 90s-long optical DNP experiment performed with a 25MHz MW bandwidth. A 60s-long pulsed SL readout was used. Solid lines indicate a Gaussian fit. Peak broadening, and the presence of a second peak at 2.72GHz indicates poor overlap of the four NV center families due to the chosen orientation. DNP at 100K is observed to be more efficient than at RT (see zoomed inset (i)).}
\zfl{mfig16new}
\end{figure}

\subsection{EPR Spectrum of NV mapped at low temperatures}
Our next focus is on cryogenic NV${\rt}\Cs$ hyperpolarization at 100K. The protocol employed has been previously described in Ref.~\cite{Ajoy17, Ajoy_DNPCombs_2018} It utilizes chirped MW pulses across the NV EPR to drive the polarization transfer. A useful initial step is mapping the NV EPR spectrum via DNP (\zfr{mfig16new}). Here, chirped MWs are applied in small windows, each spanning $\xD f{=}25$MHz, with their center frequencies sequentially swept across a broader range. This indirectly reveals the underlying EPR spectrum; following Ref.~\cite{Pillai21}, one can show that the DNP enhancement obtained is proportional to the local electronic density of states being swept over.

\zfr{mfig16new} shows the experimental data comparing the DNP-EPR spectra obtained at RT and 100K under 60s of optical pumping. Notably, we observe a shift in the EPR spectrum at lower temperatures, which is as expected from the temperature dependence of the NV zero-field splitting (ZFS)~\cite{Acosta10}. When comparing the rates of polarization transfer, there is a threefold increase in the DNP signal at 100K. We attribute this increase to the longer nuclear $T_1$ relaxation times at lower temperatures~\cite{Jarmola12}.



\begin{figure}[t]
  \centering
  {\includegraphics[width=0.49\textwidth]{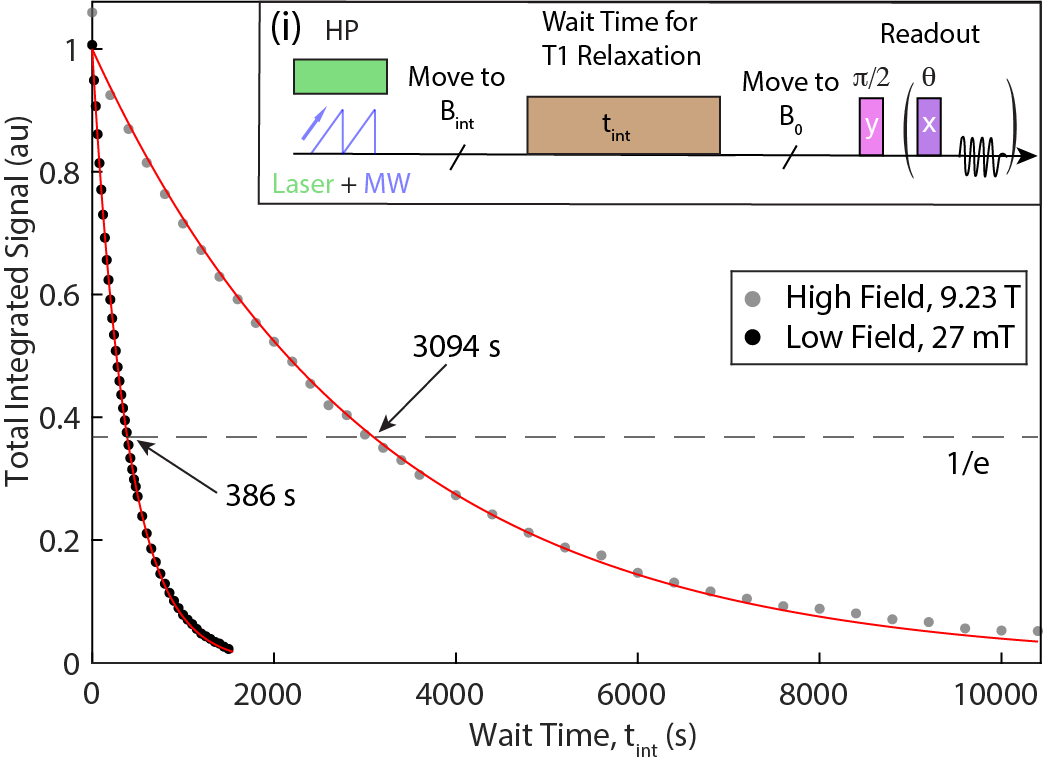}}
  \caption{\T{Field-cycling relaxometry with hyperpolarized nuclei.}
(i) \I{Experiment schematic}: Spins are allowed to relax at an intermediate field $B_\R{int}$ for a period $t_{\R{int}}$, followed by high field (9.4 T) readout via spin lock. Representative $T_1$ relaxation traces at 100K measured at low-field (27 mT) and high-field (9.4 T). Each point represents an integrated single shot measurement, taken with 60s SL readout.}
\zfl{mfig18new}
\end{figure}

\subsection{Cryogenic $T_1$ relaxometry of hyperpolarized nuclei}
The field cycling capability of our instrument, combined with the enhanced SNR of the SL sequence, enables investigation of $T_1$ relaxation times of hyperpolarized nuclei across various magnetic fields. This can be conducted by allowing hyperpolarized spins to relax at a target field, $B_{\R{int}}$, for a set period before SL readout at high field (\zfr{mfig18new}). Measuring $T_1$(B) at adjustable temperatures can identify mechanisms affecting nuclear spin relaxation. As a proof of concept, \zfr{mfig18new} shows variation of nuclear $T_1$ at low (27mT) and high magnetic field (9.4 T) at 100K. Corresponding lifetimes are recorded as $T_1$=386s at low field and $T_1$=3094s at high field. Each data point is a single-shot measurement, utilizing the high SNR from the SL readout (\zfr{mfig5}B).




\begin{figure}[b]
  \centering
  {\includegraphics[width=0.49\textwidth]{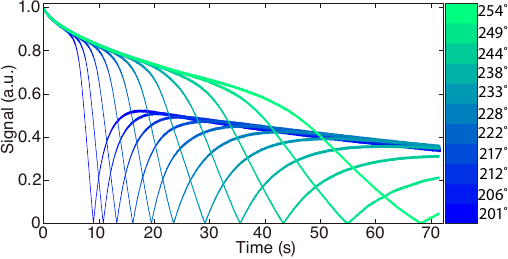}}
  \caption{\T{Creation of shell-like spin texture} with hyperpolarized $\Cs$ nuclei (Ref.~\cite{Harkins23}). This exploits a SL sequence using ${\app}\pi$-pulses creates nanoscale shells of positive and negative nuclear polarizations. Legend displays angle of pulse used in the SL sequence. Here, the pulse length $t_p{=}68 \mu$s corresponds to $\pi$. Sharp dip at zero in each of the curves corresponds to when the polarizations in the positive and negative shells are equal.}
\zfl{mfig10}
\end{figure}

\subsection{Creation of stable spin textures using electronic gradients}

As an application of the instrument, we highlight a recent experiment demonstrating the formation of stable “shells” of $^{13}$C nuclear polarization around the NV electron. A comprehensive discussion of these experiments is available in Ref.~\cite{Harkins23}.

The core concept involves leveraging the intrinsic hyperfine field from the NV electron under high magnetic fields and cryogenic conditions (e.g., 9.4 T and 100 K). Applying a CPMG pulse train consisting of $\xt{\app}\pi$ spin-lock pulses in the presence of this hyperfine field gradient causes nuclei to experience varying effective $\pi$-pulses. This results in a Hamiltonian with alternating magnetization terms across imposed domain boundaries (\zfr{mfig10}). Initially hyperpolarized nuclear spins along the $\zhat$-axis ``pre-thermalize” under this Hamiltonian~\cite{Beatrez21}, forming a nanoscale spin-textured state. This state, spanning several nanometers around the NV center, can remain stable against spin diffusion for several minutes, and incorporate hundreds of nuclear spins~\cite{Harkins23}.

A simple demonstration of this phenomenon is presented in \zfr{mfig10}. In these experiments, pulsed spin-lock measurements similar to those in \zfr{mfig17new} are performed, using ${>}5$M pulses but with a flip angle near $\xt{\app}\pi$. Unlike the smooth decays seen in \zfr{mfig17new}, the spin dynamics here exhibit sharper features, including "zero-crossings", due to a sign change in net magnetization. These zero-crossings occur where the integrated contributions of positive and negative polarized spins cancel. They can be varied with $\xt$ (as shown in \zfr{mfig10}), reflecting the ability to change the size of the texture on demand. Overall, while a fuller description is in Ref.~\cite{Harkins23}, these results demonstrate the instrument’s ability to probe hyperpolarized spin dynamics at cryogenic temperatures with high temporal precision.

\section{Connection to previous work}
In this section, we place our work in the context of prior research. Field-cycling methodologies have a storied history, beginning with Redfield’s pioneering contributions~\cite{redfield2003shuttling} and subsequently advanced by the groups of Vieth and Ivanov~\cite{grosse1999field,Kiryutin16,Zhukov18}, Ferrage~\cite{Charlier13,cousin2016high}, Pileio~\cite{Hall20}, Chou~\cite{chou2012compact,chou2016high}, and coworkers. Field-cycling relaxometry~\cite{Kimmich04,steele2016new}, in particular, has traditionally relied on shuttling methods to dynamically vary the magnetic field experienced by a sample.

Our approach builds on these foundational efforts by integrating two key advancements: the ability to maintain samples at cryogenic temperatures down to $~\sim$10 K, while performing field cycling andsimultaneous optical excitation, and the ability to perform nuclear spin manipulations involving millions of pulses, even under these cryogenic conditions.

We recognize the work of Kiryutin et al.~\cite{Kiryutin16}, who achieved cryogenic shuttling at temperatures down to 185 K. Our system extends this capability, potentially reaching temperatures $~\sim$4 K. However, we acknowledge that our shuttling speeds, while sufficient for our objectives, are slower than those reported by Ferrage~\cite{Charlier13} and Chou~\cite{chou2012compact}, which achieved sub-50 ms shuttling. This tradeoff imposes limitations on the sample types we can study but is offset by the broader temperature range and enhanced spin control capabilities our system provides.

\section{Outlook}
This instrument offers several promising avenues for further development.

\T{\I{(i) Faster low-temperature shuttling:}} Faster shuttling could be achieved by eliminating the need to move the entire cryostat. For instance, a tall cryostat volume could maintain cryogenic temperatures, while the sample is shuttled within the cryostat using a carbon-fiber rod and interfaced similar to the tuning/matching O-ring arrangement described in \zfr{mfig2}C. Alternatively, constructing the cryostat from non-metallic materials like glass or high-density plastics such as PEEK could reduce eddy current effects and allow for faster shuttling. This would broaden the range of materials that can be studied with the instrument.

\T{\I{(ii) Enhanced NMR Sensitivity:}} Signal-to-noise ratios in NMR readout could be significantly improved by integrating the first stage of the preamplifier directly within the cryostat or in the cryocooler. Similar strategies have been recently explored in the context of EPR by the Morton group~\cite{kalendra2023q}, and adapting this approach could yield valuable benefits in future iterations of this system.

\T{\I{(iii) Variable Field Hyperpolarization:}} While hyperpolarization in this work is primarily described at S-band fields, the shuttling geometry enables polarization at variable fields including at X/Q/W-bands. This would leverage the co-linearity of laser excitation with the magnetic field axis. 

\section{Conclusions} 
In conclusion, we have developed a novel instrument enabling field-cycling optical DNP across a wide magnetic field range (10 mT–9.4 T) and temperatures (10 K–RT). This system seamlessly integrates low-field hyperpolarization with high-sensitivity NMR detection at high fields, while accommodating variable cryogenic conditions. The instrument opens exciting experimental opportunities, including the optimization of hyperpolarization in emerging platforms, particularly molecular systems, and variable-field relaxometry to probe the underlying mechanisms of relaxation processes in these systems.

\section*{Acknowledgements} 
We gratefully acknowledge discussions with J. Reimer, D. Suter, and C. Ramanathan and technical assistance from J. Mercade and M. Elo (Tabor Electronics). This work was funded by AFOSR DURIP (FA9550-22-1-0156), NSF MRI (2320520), ONR (N00014-20-1-2806), AFOSR YIP (FA9550-23-1-0106), NSF TAQS (2326838), and the CIFAR Azrieli Foundation (GS23-013). KAH acknowledges an NSF Graduate Research Fellowship. CS acknowledges an ARCS Graduate Fellowship.

\paragraph*{\textbf{Author Contributions:}}

ND, KAH, and DM contributed equally to this work. AA and DM conceptualized the instrumentation and DM, ED, and PR designed the instrumentation. ED, AN, MH visualized the instrument. DM, KAH, ND, CS, ED, ZZ, and PR constructed the instrument. KAH, ND, and DM implemented software for instrument control and ND and KAH performed debugging. KAH, ND, CS, and SB collected data; and ND, KAH, and CS processed data. ND, KAH, SB, UB, and ZZ all visualized data. AA, ND, and KAH drafted the initial manuscript with manuscript editing by KAH, ND, AA, UB, and SB. Data storage and curation was performed by MM.

\vspace{-5mm} 
\vspace{-1mm}

\end{document}